Main Articles

# Fast-Forward on the Green Road to Open Access: The Case Against Mixing Up Green and Gold

**Stevan Harnad** provides a summary of his critique of Jean-Claude Guédon's article on the green and gold roads to Open Access.

Main Contents | Section Menu | Email Ariadne | Search Ariadne

## Introduction

This article is a critique of: The "Green" and "Gold" Roads to Open Access: The Case for Mixing and Matching by Jean-Claude Guédon [1].

Open Access (OA) means: free online access to all peer-reviewed journal articles.

Jean-Claude Guédon (J-CG) argues against the efficacy of author self-archiving of peer-reviewed journal articles -- the "Green" road to OA -- on the grounds (1) that far too few authors self-archive, (2) that self-archiving can only generate incomplete and inconvenient access, and (3) that maximizing access and impact is the wrong reason for seeking OA (and only favours elite authors). J-CG suggests instead that the right reason for seeking OA is so as to reform both the peer review system and the journal publishing system by creating new ways of "creating value" and "branding" whilst converting to OA ("Gold") publishing (in which the online version of all articles is free to all users). We should convert to Gold by "mixing and matching" Green and Gold as follows:

First, self-archive dissertations (not published, peer-reviewed journal articles). Second, identify and tag how those dissertations have been evaluated and reviewed. Third, self-archive unrefereed preprints (not published, peer-reviewed journal articles). Fourth, develop new mechanisms for evaluating and reviewing those unrefereed preprints, at multiple levels. The result will be OA Publishing (Gold).

I reply that this is not mixing and matching but merely imagining: a rather vague conjecture about how to convert to 100% Gold, involving no real Green at all along the way, because Green is the self-archiving of *published, peer-reviewed articles*, not just dissertations and preprints.

I argue that rather than yet another 10 years of speculation [2] what is actually needed (and imminent) is for OA self-archiving to be mandated by research funders and institutions so that the self-archiving of *published, peer-reviewed journal articles* (Green) can be fast-forwarded to 100% OA. The direct purpose of OA is to maximise research access and impact, not to reform peer review or journal publishing; and OA's direct benefits are not just for elite authors but for all researchers, for their institutions, for their funders, for the tax-payers who fund their funders, and for the progress and productivity of research itself.

There is a complementarity between the Green and Gold strategies for reaching 100% OA today, just as there is a complementarity between access to the OA and non-OA versions of the same non-OA articles today. Whether 100% Green OA will or will not eventually lead to 100% Gold, however, is a hypothetical question that is best deferred until we have first reached 100% OA, which is a direct, practical, reachable and far more urgent immediate goal - and the optimal, inevitable and natural outcome for research in the PostGutenberg Galaxy.

"Recent discussions on Open Access (OA) have tended to treat OA journals and self-archiving as two distinct routes"

From the day it was coined in 2001 by the Budapest Open Access Initiative (BOAI), [3] "Open Access" has always been defined as free online access, reachable by two distinct routes, BOAI-1, OA self-archiving ("Green") and BOAI-2, OA journals ("Gold"):

"Some... even suggest that [self-archiving] alone can bring about full Open Access to the world's scientific literature"

OA's focus is only on peer-reviewed journal articles -- 2.5 million annual articles in 24,000 peer-reviewed scholarly and scientific journals -- not on "the world's scientific literature" in its entirety (i.e., not on books, magazines).

(1) To self-archive one's own article is to provide Open Access (OA) to one's own article. Every author can do this, for every one of his articles. If/when every author does this, for each of the annual 2.5 million articles, we have, by definition, 'full Open Access' (Green).

(2) By the same token, if/when every publisher of each of the 24,000 journals converts to OA publishing, we have, by definition, 'full Open Access' (Gold).

The rest is simply a question of probability: Is it more probable that all or most journals will convert to OA, or that all or most of their authors will self-archive their articles? Which faces more obstacles, costs, delay, uncertainty, risk? Which requires more steps? Which can be facilitated by university and research-funder OA mandates? Which is already within immediate reach?

"[S]elf-archiving is not enough... the repositories [need] some branding ability"

Self-archiving is not enough for what? Would 100% self-archiving not correspond to 100% OA (just as 100% OA journals would)?

And as we are talking about the self-archiving of peer-reviewed, published journal articles, why is there a need for "branding"? Branding what? The journal articles? But those are *already* branded -- with the name of the journal that published them. What is missing and needed is not *branding* but *Open Access* to those journal articles! (J-CG's preoccupation with branding will turn out to be a consequence of the fact that he is not proposing a way to make current journal articles OA, but a way to replace current journals altogether.)

"[OA journal publishing] amounts to a reform of the existing publication system [relying] on journals as its basic unit... and... aims at converting [to] or creating... Open Access journals."

*Both* OA self-archiving and OA journal publishing (and indeed, OA itself, and the definition of OA) "fundamentally rely on journals as [their] basic unit" because it is the articles in peer-reviewed journals that are the target literature of the OA movement.

It is true, however, that only BOAI-2, OA journal publication (Gold), aims at a reform of the existing publication system. BOAI-1, OA self-archiving (Green), is neutral about that. It aims only at OA...

"[F]inancial viability [of OA publishing] rests on the will of governments ... and varies... with... country and circumstances"

All the new and converted OA journals are valuable and welcome, but their numbers and the rate of increase of their numbers has to be realistically noted: About 5% of journals are OA ("gold") journals today (1400/24,000). In contrast, about 92% of journals are "green" - i.e., they have given their authors the green light to self-archive their articles if they wish. The rate of increase in the number of green journals has been incomparably faster than the rate of increase in the number of gold journals in the past few years. The amount of OA (15%) generated via self-archiving has also been three times as great as the amount of OA generated via OA publishing (5%); and (although direct measures have not yet been made) it is likely that the rate of growth of OA via OA self-archiving is also considerably higher than the rate of growth of OA via OA publishing - for obvious reasons that have already been mentioned: It is far easier and cheaper to create and fill an institutional OA Archive than to create and fill an OA journal. Moreover, there is a considerable financial risk for an established journal in converting to the OA cost-recovery model, which has not yet been tested long enough to know whether it is sustainable and scaleable.

So whereas all new and converted OA journals are welcome, it makes no sense to keep waiting for or focusing on them as the main source of OA. The real under-utilised resource is OA self-archiving - underutilised even though it already provides three times as much OA as OA journals and is probably growing faster too: because OA self-archiving is already in a position to provide immediate 100% OA, if only it is given more of our time, attention and energy.

It is unrealistic to imagine that the reason the number of new and converted gold journals is not growing faster is that governments are not willing to subsidise them! It is not clear whether governments should even want to, at this point, when OA is already reachable without any need for subsidy, via self-archiving, and researchers are simply not yet ready to perform the few keystrokes required to reach for it (even for the 92% of their articles published in green journals) despite being alleged to want and need OA, despite being willing (in their tens of thousands!) to perform the keystrokes to demand it from their publishers -- despite the fact that the benefits of OA itself are intended mainly for researchers and research [4].

If government intervention is needed on behalf of OA, surely it is needed in order to induce their researchers to provide the OA that is already within their reach to provide, rather than to subsidise journals to do it for them.

"[For self-archiving] the article [is the] fundamental unit [and] journals matter only to differentiate between peer-reviewed articles and non-peer-reviewed publications and to provide symbolic value"

**created using Corda Builder**

*Symbolic value*? Consider how much simpler and more straightforward it is to state this theory-independently: Today, most of the 2.5 million articles published in the world's 24,000 peer reviewed journals are inaccessible to many of their potential users because they cannot afford access. If the articles are made accessible free online (by self-archiving them), this problem is solved.

We need not theorise about *why* users want to use peer-reviewed journal articles. We can take that as a rather obvious premise. Yes, users want the peer-reviewed articles (and the journal's name tells them which ones those are); and peer review itself provides the 'value' they seek in an expert-vetted literature rather than an unfiltered free-for-all. There is no need to debate the value of peer review in an OA context: One of the *premises* of the OA movement is that OA is about access to the peer-reviewed journal literature, not access to something else. So peer review and the journal names come with the territory. The only problem to solve is *access*. And Open Access solves that. And self-archiving is by far the fastest and surest way to provide immediate OA.

No further theorizing, or complicating, is needed: We have peer-reviewed journal articles, but we do not have Open Access to them. Self-archiving them provides that access. End of story. The rest is merely speculation (needless speculation, needless complication), needlessly delaying OA.

"The problem with the self-archiving argument is that, until now at least, its results are unimpressive"

OA self-archiving's results to date are indeed unimpressive. I do not disagree at all -- *but compared to what*? Certainly not compared to the results of *OA publishing*, since OA self-archiving has generated 3 times as much OA as has OA publishing and is probably growing much faster too. Green is only unimpressive *relative to its own immediate potential for generating OA*, which is at least 92%, compared to Gold's 5%. In that respect, it can be said that Gold, at 5%, is much closer to its full immediate growth potential, whereas Green, at 15%, is not. But surely the remedy for that is to devote more time, attention and energy to exploiting Green's full immediate potential! That is what the impending self-archiving mandates will do. In the meanwhile, however, it would help if (1) less time and attention were devoted exclusively to Gold, as if OA publishing and OA were the same thing, and if (2) the Gold option were always balanced by pointing out the Green option too.

(Green has for several years now adopted the unified OA provision strategy: 'If there is a suitable Gold journal for your paper, publish it there; if not, publish it in a Green journal and self-archive it.' Just taking that step of fairly presenting the two options at all times would go a long way toward redressing the imbalance between Gold and Green.)

"More important still, a majority of scientists have not been swayed either"

Not just *more* important: *most* important. Indeed the problem of "swaying" researchers to provide the OA that they are purported to want and need so much is the *only* real challenge for OA. And it is already clear what will meet that challenge: (1) Empirical evidence of the OA impact advantage [5] plus (2) an OA self-archiving mandate on the part of researchers' institutions and research funders to ensure that advantage is taken of that advantage - by naturally extending their existing 'publish or perish' mandate to 'publish *and* self-archive' (so as to maximise the access to, and the usage and impact of, your articles) [6].

And we already know from a recent survey that just as they currently comply with their 'publish or perish' mandate, most researchers report they will *not* self-archive if it is not mandated, but they *will* self-archive - and self-archive *willingly* -- if ever it is mandated by their institutions or funders [7].

"the number of articles published in "gold" journals (5%)... is often contrasted with the total number of articles published under "green" titles (85% or more), without any mention.. that a majority of those are not actually...in Open Access repositories"

On the contrary, it is *always stated very explicitly* (including in an article co-appearing in the very same issue as J-CG's article!) [8] that whereas 93% of journals are green, only 20% of articles are OA, 15% of them OA via self-archiving! This fact is not being concealed, it is being *trumpeted*, in order to point out that if researchers really want and need OA and its benefits in terms of access and impact as much as they are described (by OA advocates of both the Green and Gold hue) as wanting and needing it, then it is up to researchers to provide it - particularly where they have even been given their publisher's green light to go ahead and do so!

But it is clear that just as far fewer researchers would publish anything at all (despite the advantages of publishing -- advantages that researchers presumably want and need) if it were not for their institutions' and research funders' "publish or perish" mandate to do so, so researchers will likewise not self-archive until their institutions and research funders make their employment, salary and research funding conditional on their doing so. (Institutions and funders already do this implicitly, in making researchers' employment, salary and research funding conditional not only on publication, but on the *impact* of publication. Since OA maximises impact, this implicit causal connection and contingency now simply needs to be formalised explicitly.)

"Institutional archives are being created, but need to be filled more quickly, by authors, with research journal papers. Attracting authors and their papers requires evidence of services that will improve the visibility and impact of their works"

Correct, and we are gathering and disseminating the requisite evidence [9]. However, as noted, this evidence, and the probability of enhanced usage and impact to which they attest, are still not enough to induce a high or fast enough rate of OA-provision, just as the probability of the usage and impact that will result from publishing at all are not enough to induce publishing: The incentive to provide OA, like the incentive to publish, must be made explicit, as being among the formal conditions for employment, promotion and research funding, much the way both publishing and research impact are already among the conditions for employment, promotion and research funding.

"No wonder... scientists are not rushing to self-archive; no wonder... the "self-archiving" side has welcomed mandating "self-archive" ... If research institutions... through their promotion and tenure procedures, and the granting agencies, through their evaluation procedures, favor documents in Open Access.. then Open Access will indeed progress. But...[this] argument [is] totally independent [of] the impact advantage argument"

*Institutions and funders mandating OA for their research output has nothing to do with the "impact advantage argument" (and evidence)*? One might as well say that the existing weight that institutions and funders place on journal impact factors has nothing to do with impact either!

The very *reason* for OA itself, and the institutional and funder *rationale* for mandating that OA should be provided (by self-archiving) has *everything* to do with impact. It is in order to maximise the visibility, usage impact and citation impact of their research output -- instead of continuing to limit it to only those users whose institutions can afford to pay for access -- that universities and granting agencies are now planning to mandate self-archiving:

(Does J-CG imagine, instead, that they are mandating OA in order to reform the publishing system or to solve the journal pricing/affordability problem?)

"Where governments decide to move in and press for Open Access publications, a great deal of... political groundwork [is needed]. But... the need to rely on mandating shows that the "self-archiving" side cannot avoid political maneuvers either"

Governments cannot and do not "move in and press for Open Access publications": Whom can they press, how, to do what? All that governments can do is to cover OA journal authors' publication fees. What they can press for is *Open Access itself*. Research funders can require their grantees to *provide OA* to their funded research findings -- either by self-archiving all the resulting non-OA journals articles (Green) *or* (if a suitable journal exists) by publishing them instead in an OA journal (Gold) -- as a precondition for receiving research funding at all. And research funders now seem well on the way to pressing for exactly that. If/when they do, we will all be well on the way to 100% OA [10].

"Ultimately, the point of the "gold" road is to create intellectual value in new and better ways. To achieve this goal, the "gold" road must pay attention to more than the impact advantage that addresses only the author side of the scientist. A scientist is also a hurried reader and value can be built out of better searching, retrieving, and navigating tools... The "gold" projects should strive to collaborate to create citation links and indices"

"Create intellectual value"? Is it not simpler and more theory-neutral to say that OA journals make their articles OA, and charge the author-institution, whereas non-OA journals charge the user-institution? Do we really need the theorizing about "creating intellectual value"? After all, we all know what journals do: they implement peer review and tag the result with the journal's name, and its associated track-record for quality. OA and non-OA journals both do that; they just charge different parties for it.

Gold journals should certainly collaborate and citation-link, and they do. But so do all the non-gold journals (green and grey) do the same. And so do OA Archives and harvesters, like Citebase [11]. So what is the point here?

"I would suggest starting not with peer-reviewed articles, but rather with doctoral dissertations [emphasis added]"

Here we have arrived at the heart of J-CG's proposal. OA's target is peer-reviewed articles. Only 20% of them are OA to date. And J-CG suggests that we not "start" with them, but with self-archiving dissertations (some of which are being self-archived already) [12]. Why are we being asked to start our mixing match with dissertations?

(The answer will shortly become apparent: We will *never* in fact be moving on to self-archiving peer-reviewed journal articles, in J-CG's recommended mix/match scenario: We will instead somehow - it is never specified quite how (but one guesses that it is by going on to self-archive unrefereed preprints rather than peer-reviewed journal articles) - *segue* into the construction, bottom-up, of a brand new alternative peer review and publication system, somehow, on top of these dissertations, to replace the existing peer-reviewed journals. This is in fact J-CG's version of the Golden road, of creating OA journals; it is not a mix-and-match of green and gold at all! It is an independent reinvention of the entire wheel. The first step is new forms of "quality evaluation" -- arising, somehow, out of new ways of "promoting the intellectual value" of doctoral dissertations...)

"an inter-institutional strategy to promote the intellectual value, authority, and prestige of doctoral theses could easily provide the testing ground for the emergence of inter-institutional disciplinary archives"

At a time when out of 2.5 million annual peer-reviewed articles in 24,000 journals -- all of them already having whatever intellectual value they already have  -- only 20% are OA, we are being bidden to self-archive dissertations and to promote their intellectual value! But is not it the other 80% of those already evaluated but still inaccessible articles that we really need? And is this, then, the promised alterative green-and-gold mix-and-match proposal?

"Evaluation Levels: The metadata should also be extended to provide some indication of quality. It could be designed to help identify the identity and the nature of the evaluating body that passes judgment over the documents in the repository. In other words, the metadata should help identify the quality, nature, and procedures of groups that begin to work as editorial boards would. The metadata could also help design evaluations [sic] scales - imagine a one brain, two brain, ... n-brain scale, similar to a Michelin guide for restaurants"

Quality metadata? For dissertations? Whatever for? We are not trying to create a dissertation peer-review service. We are trying to provide OA for the other 80% of already peer-reviewed articles; and their "indication of quality" is the name of the journal that already performed the peer review on them!

For journal articles, the "evaluating body" is the journal, tagged by its name. For dissertations, who knows? (And who cares, insofar as OA is concerned? Quality-tagging dissertations has nothing to do with OA.)

We do not need a Michelin guide to journal articles! We just need access to the articles, and their journal names! Here we are instead busy reinventing peer review, in hitherto untested and (as we shall see) rather extravagant new forms:

"This leads to a new project: If various universities create consortia of disciplinary repositories, then nothing prevents them from designing procedures to create various levels of peer review evaluation, e.g., institutional, consortial, regional, national, international"

This is a new project indeed! For the old project was to provide Open Access to the remaining 80% of the existing 2.5 million annual articles in the existing 24,000 peer-reviewed journals. Now we are talking about what? Reviewing the peer review? Re-doing the peer review? Replacing the peer review?

I suspect that - driven by his publication-reform theory - what J-CG is really contemplating here is entirely replacing the non-OA journals in which 95% of these self-archived articles are currently being published! But OA itself is just about providing free, online access to those peer-reviewed journal articles, not about wresting them from the journals that published them, and from their peer-review - at least not OA via the Green road of self-archiving.

Nor is that what a journal means in giving its green light to self-archiving: that you may treat my published articles as if they had never been peer-reviewed and published, and simply start the whole process anew! Nor is there any point in starting the whole process anew, if it was done properly the first time. It is hard enough to get qualified referees to review papers once, let alone to redo it (many times? at many "levels"?) all over again.

Nor is that what the authors or the would-be users of those already peer-reviewed, published articles need or want. What authors and users want is Open Access to those articles.

How did we get into this bizarre situation? It was by accepting the invitation to populate the OA Archives with dissertations instead of peer-reviewed journal articles, as a "testing ground." Apparently, we are never to graduate to self-archiving journal articles (Green) at all: Rather, we are to rebuild the whole publication system from the bottom up, starting with dissertations, and then generalizing and applying it to the unrefereed preprints of articles-to-be. (J-CG has wrongly inferred or imputed that "Green" means mainly the self-archiving of unrefereed preprints, rather than the peer-reviewed, published postprints that are OA's main target!)

In other words, we are to reform the publication system after all, just as J-CG recommended. Never mind about providing Open Access to that other 80% of existing journal articles: We will instead create new evaluation bodies (Gold). We have time, and surely all parties will eventually go along with this project (in particular, all those sluggish authors who had not been willing even to self-archive...). They are to be weaned from their current journals and redirected instead to - it is not yet altogether clear what, but apparently something along the lines of: "various levels of peer review evaluation, e.g., institutional, consortial, regional, national, international"...

"At that point, a recognised hierarchy of evaluation levels can begin to emerge [emphasis added]... clearly identifiable through the metadata... [indicating] what level of peer review and evaluation is being used... [and] which group is backing it. In effect, this is what a journal does and... how it acquires some branding ability"

In effect, the current journal system is what J-CG is proposing (because 95% of journals have obstinately declined to go Gold) to *replace* (bottom-up, starting with dissertations) with the above "emergent" Gold system, consisting of "a recognised hierarchy of evaluation levels" ("clearly identifiable through the metadata"). And this alternate "branding" system, starting with dissertations, is to emerge as a result of mixing-and-matching green-and-gold.

What follows is speculation piled upon speculation, all grounded in this initial premise (that if you cannot convert them, replace them):

"An international registry of such evaluation procedures and of the teams of scholars involved could then be developed... [to] lend transparency and credibility to these value-building procedures. In this fashion, a relatively orderly framework for expanded peer review and evaluation can emerge"

J-CG has **here** managed to resurrect, almost exactly, the very same incoherence that beset Harold Varmus's original 1999 E-biomed proposal, which could never quite decide whether what was being proposed was: (1) free access to journal articles, (2) an archive in which to self-archive articles to make them freely accessible, (3) a rival publisher or publishers to lure away authors from existing journals (4) an alternative kind of journal or journals, with alternative kinds of peer review, or (5) all of the foregoing [13].

That somewhat mixed-up and ill-matched 1999 vision is what has since become gradually more sorted out and focused in the ensuing years, as follows: First, PubMed Central [14] was created (February, 2000) as a central OA archive in which publishers were invited to deposit their contents six months after publication. When few publishers took up that invitation, the Public Library of Science (PloS) was founded (October 2000) and circulated an Open Letter, signed by 34,000 biomedical researchers the world over, demanding that existing journals should go Gold [15]. When that too failed, PLoS became an Open Access Publisher (2001) and has since launched two Gold journals [16] (forgetting altogether about green self-archiving). Most recently (2004), perhaps having noticed that the golden road to OA is a rather long and slow one, PLoS again took up the green road of self-archiving by helping to promote the proposed NIH public-access policy (which would request that all articles resulting from NIH-funded research should be self-archived in PubMed Central within 6 months of publication) [17].

J-CG now proposes to resurrect something very much like the original 1999 matchless E-biomed mix once again! I would like to make the counter-proposal that once was enough: that we should forget about trying to rebuild the publishing system bottom up (with a vague, untested, speculative, and probably incoherent mix-and-match model of archiving and publishing) and instead reinforce the road that Harold Varmus, PLoS, the Wellcome Trust, the UK Government Select Committee, and many others have lately rejoined (and the one they would have been better off taking in the first place) -- the green road to OA -- by promoting the OA self-archiving of the remaining 80% of the existing journal literature as a condition of research funding (and employment).

"New Journal Models: Transparency, prestige, and rigor are needed to create credible value... something like "overlay journals"[will] begin to emerge and... gradually acquire visibility and respect. At that point, the institutional repositories will have effectively morphed and matured into a consortium-based network of repositories with a rich set of value-creation tools and increasingly recognised names or labels"

The trouble is that all the "morphing" so far is happening only in the mind of the passive speculator; and meanwhile 80% of articles continue to be inaccessible to those would-be users who cannot afford access - yet *that* was supposed to be the problem OA was remedying.

Just like E-biomed and "deconstructed journals," "overlay journals" are at the moment figments of the armchair theorist's imagination. Moreover, it is not even clear what "overlay journals" means. If it just means conventional journals (whether hybrid or online-only) implementing online peer review by having submissions deposited on a Web site and then directing referees and revised drafts to the site, then most journals are already overlay journals in this banal sense.

If "overlay journals" means journals that are online only, then that is nothing new or interesting either. If it means that the archive to which the referees go to find the paper and where revised drafts are put is not the journal's Web site but an OA Archive (whether institutional or central), then that too is uninteresting -- just a trivial (and quite natural) implementational variant of a standard feature of extant journals and conventional (online) peer review [18].

If the journal itself performs *only* peer review and certification, and the archives do all access-provision and archiving, then this may have some potential interest, some day -- but there exist at most a handful of journals that resemble that description today, and between them and the remaining 99.99% of journals is the still unsettled future of OA (Gold) journals and their cost-recovery model [19].

So overlay journals are still just an armchair speculation: But 100% OA need not be - and it certainly need not wait for the "morphing" of the current 24,000 journals into overlay journals.

"As a result of this evolution... original submissions will be addressed to these new channels of scientific communication [overlay journals]... in parallel with, the fact [they] will have been already "published" in traditional journals. This is where the importance of "self-archiving" really finds its anchoring point... All this can be achieved by treating the "self-archiving" strategy as a transition phase on the way to the "gold" objective"

It is hard to see how an article that *has already been published in a traditional journal* can become an "original submission" to an "overlay journal" - harder still to see this as self-archiving's real "anchoring point." Perhaps self-archiving should just stick to the more mundane task of providing immediate OA to the remaining 80% of the current journal literature, rather than waiting for this hypothetical new multilevel, multivalent system to evolve?

The only "transition phase" that is worth talking about (and is tested, and visible, and reachable) is the transition from today's 20% OA to 100% OA via self-archiving. After that, *nolo contendere* - and *hypotheses non fingo*!

"Open Access should not be the tactical tool of a few, elite, established, scientists that want to enhance their careers and little else"

No one has suggested OA is, or should be the tactical tool of a few, elite, established, scientists. It is J-CG, however, who suggested (without saying how, or why) that impact enhancement through OA self-archiving would only benefit the elite, established scientists. The analysis by author/article seniority and quality-level is yet to be done, but there is no particular reason to expect that the OA-impact advantage will be only, or even mostly, at the top.

## Conclusion

What is needed today is already quite clear: 100% OA by the fastest and surest means possible. It is also clear what that means is: self-archiving (Green), which now needs to be mandated by researchers' institutions and funders. There is also scope both for the growth of OA journals (Gold) and for experimentation with hypothetical new systems *in parallel* with the self-archiving of all peer-reviewed, published journal articles (Green) - but not *in place* of it. Let there be no mix-up about that!

# References

1. All quotes are from: *Serials Review* 30(4) 2004 http://dx.doi.org/10.1016/j.serrev.2004.09.005 For a fuller version of this critique, see: http://www.ecs.soton.ac.uk/~harnad/Temp/mixcrit.html ; Editor's note: some useful information on *Serials Review* 30(4) 2004 is available at http://www.library.yale.edu/~llicense/ListArchives/0412/msg00075.html
2. Richard Poynder, "Ten Years After" *Information Today* 29(9) 2004. http://www.infotoday.com/IT/oct04/poynder.shtml
3. Budapest Open Access Initiative. http://www.soros.org/openaccess/read.shtml
4. A Keystroke Koan for our Open Access Times. http://www.ecs.soton.ac.uk/~harnad/Hypermail/Amsci/3061.html
   Journal Self-Archiving Policy Registry. http://romeo.eprints.org/stats.php
5. Bibliography of Open Access Impact Advantage Studies. http://opcit.eprints.org/oacitation-biblio.html
6. Registry of Institutional Open Access Provision Policies. http://www.eprints.org/signup/sign.php
7. Swan, A & Brown, S. Authors and open access publishing. *Learned Publishing* 17 (3) 2005 http://www.keyperspectives.co.uk/OpenAccessArchive/Authors_and_open_access_publishing.pdf
8. Harnad, S., Brody, T., Vallieres, F., Carr, L., Hitchcock, S., Gingras, Y, Oppenheim, C., Stamerjohanns, H., & Hilf, E. (2004) The Access/Impact Problem and the Green and Gold Roads to Open Access, Serials Review 30 (4) 2004 http://dx.doi.org/10.1016/j.serrev.2004.09.013
9. Open Access Impact Studies: http://citebase.eprints.org/isi_study/ http://www.crsc.uqam.ca/lab/chawki/ch.htm
10. Directory of Institutions with Open Access Policies. http://www.eprints.org/signup/fulllist.php
11. Citebase scientometric search engine http://citebase.eprints.org/
12. Open Access Dissertation Archives. http://archives.eprints.org/eprints.php?page=all&type=theses
13. Critique of e-Biomed Proposal. http://www.nih.gov/about/director/ebiomed/com0509.htm#harn45
14. PubMed Central. http://www.pubmedcentral.nih.gov/
15. PLoS Open Letter to Publishers. http://www.plos.org/support/openletter.shtml
16. PLoS Journals. http://www.plos.org/journals/
17. Proposed NIH Public Access Policy, http://grants.nih.gov/grants/guide/notice-files/NOT-OD-04-064.html
18. Online Peer Review Innovations. http://www.ecs.soton.ac.uk/~harnad/Temp/peerev.ppt
19. Harnad, S. "For Whom the Gate Tolls?" http://www.ecs.soton.ac.uk/~harnad/Tp/resolution.htm#4.2

# Author Details

**Stevan Harnad**
Canada Research Chair
Université du Québec à Montréal

Email: mailto:harnad@uqam.ca
Web site: http://www.ecs.soton.ac.uk/~harnad/

*Ariadne* is published every three months by UKOLN. UKOLN is funded by MLA the Museums, Libraries and Archives Council, the Joint Information Systems Committee (JISC) of the Higher Education Funding Councils, as well as by project funding from the JISC and the European Union. UKOLN also receives support from the University of Bath where it is based. Material referred to on this page is copyright Ariadne (University of Bath) and original authors.

**created using Corda Builder**